\documentclass[%
 reprint,
nofootinbib,
 amsmath,amssymb,
 aps,
 prd,
]{revtex4-2}

\usepackage{graphicx}
\usepackage{dcolumn}
\usepackage{bm}
\usepackage{hyperref}

\usepackage{subcaption}
\usepackage{anyfontsize}
\usepackage{physics}
\begin{document}

\preprint{APS/123-QED}

\title{Newtonian restricted three-body gravitational problem \\ with positive and negative point masses}

\author{Kok Hong Thong$^{1, 2, 3}$}
\email{kokhongt@student.unimelb.edu.au}
\author{Andrew Melatos$^{1, 2}$}%
\email{amelatos@unimelb.edu.au}
\affiliation{$^{1}$School of Physics, University of Melbourne, Parkville, Victoria, 3010 Australia}
\affiliation{$^{2}$ARC Centre of Excellence for Gravitational Wave Discovery (OzGrav), \\
 Hawthorn, Victoria 3122, Australia}
 \affiliation{$^{3}$Department of Applied Mathematics and Theoretical Physics, University of Cambridge, Centre for Mathematical Sciences, Wilberforce Rd, Cambridge CB3 0WA}




\date{\today}

\begin{abstract}
The Newtonian restricted three-body problem involving a positive primary point mass, $m_+$, and a negative secondary point mass, $m_-$, in a circular orbit, and a positive or negative tertiary point mass, $m_3$, with $m_+ > |m_-| \gg |m_3|$, is solved. 
Five Lagrange points are found for $m_3$, three of which are coplanar with $m_+$ and $m_-$, and two of which are not, a subtle consequence of the gravitational repulsion from $m_-$. 
All Lagrange points are linearly unstable, except for one point in the regime $m_+ \gtrsim 8.4 |m_-|$, which is linearly stable and collinear with $m_+$ and $m_-$.
\end{abstract}

\maketitle


\section{\label{sec:1}Introduction}

Bodies with negative mass are purely hypothetical; there is no experimental evidence that they exist. 
However, they remain a subject of theoretical interest. 
Negative mass has been proposed to give rise to the apparent dark matter and energy in $\Lambda$CDM-cosmology \citep{farnes_unifying_2018,socas-navarro_can_2019,najera_negative_2021}.
Negative mass is found in exotic exact solutions to Einstein's field equations, such as traversable wormholes \citep{morris_wormholes_1988}, warp drives \citep{alcubierre_warp_1994,lobo_exotic_2007}, gravitationally repulsive stars \citep{novikov_stars_2018}, and the gravitational two-body problem \citep{bondi_negative_1957,bonnor_exact_1964,podolsky_null_2001,shatskiy_kepler_2011,ivanov_relativistic_2012}. 
Negative mass has been used in the Newtonian gravitational N-body problem to either find solutions to particular configurations \citep{roberts_continuum_1999,celli_homographic_2005,celli_central_2007,manfredi_cosmological_2018,farnes_unifying_2018,rahman_existence_2019} or as a fictitious device to simplify an analysis \citep{banfi_search_1991}.
Observational signatures, such as the gravitational bending of light by negative-mass lenses, have also been studied \citep{safonova_microlensing_2001,izumi_gravitational_2013,kitamura_microlensed_2014}.

The gravitational dynamics of bodies with negative masses are counter-intuitive. 
\citet{bondi_negative_1957} found an exact solution in general relativity of the gravitational two-body problem involving one positive and one negative point mass. 
The positive mass attracts the negative mass; that is, the negative mass accelerates\footnote{Acceleration here is gravitational and is interpreted in the Newtonian framework. In general relativity, both point masses move along geodesics \citep{bonnor_exact_1964}.} towards the positive mass. 
The negative mass repels the positive mass; that is, the positive mass accelerates away from the negative mass. 
In the Newtonian regime, if the point masses have the same magnitudes and begin with the same velocities, they accelerate constantly forever, in a so-called runaway motion \citep{bondi_negative_1957,bonnor_exact_1964,bonnor_negative_1989,ivanov_relativistic_2012}. 
Although this scenario is counter-intuitive, one can verify that local mass and momentum conservation laws are satisfied. 

The three-body problem involving both positive and negative masses has not been studied in full generality. 
\citet{celli_homographic_2005} studied homographic\footnote{Homographic solutions occur when there exists a time-dependent linear similarity with proportionality constant $s(t)$ satisfying \citep{celli_homographic_2005}
\begin{equation}
    \mathbf{X}_j(t) - \mathbf{X}_i(t) = s(t) [\mathbf{X}_j(0) - \mathbf{X}_i(0)],
\end{equation} 
with $i, j \in \{1, 2, 3\}$, where $\mathbf{X}_i(t)$ is the position vector of the $i$-th body.} solutions to the three-body problem with both positive and negative masses, and a vanishing total mass.
However, the literature does not contain a full analysis of the location and stability of the stationary Lagrange points, for example.
In this paper, we study for the first time the Newtonian restricted three-body gravitational problem with a positive primary point mass, $m_+$, a negative secondary point mass, $m_-$ and a positive or negative tertiary point mass, $m_3$.
Henceforth, for brevity, we call this situation the restricted exotic three-body problem.
The masses are ordered according to $m_+ > |m_-| \gg |m_3|$. 
\citet{shatskiy_kepler_2011} performed a Newtonian analysis of the Bondi two-body problem involving arbitrary initial conditions and found elliptical, parabolic and hyperbolic solutions. 
In Section \ref{sec:2}, we review the elliptical solutions for $m_+ > 0$ and $m_- < 0$, as well as the conditions under which circular orbits occur. 
We also write down the equation of motion for $m_3$.
In Section \ref{sec:3}, we find the location of the Lagrange points.
We then study the linear stability of the Lagrange points in Section \ref{sec:4}.

\section{\label{sec:2}Restricted exotic three-body problem}
In this section, we set up the restricted exotic three-body problem. 
We present the motion of $m_+$ and $m_-$ in Section \ref{sec:2.1}.
We then discuss the equation of motion of $m_3$ in Section \ref{sec:2.2}.

\subsection{Primary and secondary motion}
\label{sec:2.1}
The Lagrangian of the primary and secondary point masses excluding the influence of $m_3$ is given by
\begin{equation}
\label{eq:Kepler}
    L_{+,-} = \frac{1}{2} m_+ \dot{\mathbf{x}}_+^2  + \frac{1}{2} m_-\dot{\mathbf{x}}_-^2 + \frac{Gm_+m_-}{\abs{\mathbf{\mathbf{x}_+ - \mathbf{x}_-}}},
\end{equation}
where $\mathbf{x}_+$ and $ \mathbf{x}_-$ are the position vectors of $m_+$ and $m_-$, respectively, and the overdot denotes a time derivative.
Equation (\ref{eq:Kepler}) defines the Kepler problem \citep{shatskiy_kepler_2011}.
We assume the equivalence principle; inertial and gravitational masses are equal in magnitude and sign\footnote{For a study of the restricted three-body problem with unequal but positive inertial and gravitational masses, the reader is referred to \citet{nordtvedt_equivalence_1968}.}.
The barycenter is located at
\begin{equation}
    \mathbf{R} = \frac{m_+ \mathbf{x}_+ + m_- \mathbf{x}_-}{m_+ + m_-},
\end{equation}
and the reduced mass is given by 
\begin{equation}
    \mu = \frac{m_+m_-}{m_++m_-}.
\end{equation}
One has $\mu > 0$ for $m_+ + m_- < 0$, and $\mu < 0$ for $m_+ + m_- > 0$, the case considered primarily in this paper.

We perform a change of variables from $\mathbf{x}_+$ and $\mathbf{x}_-$ to $\mathbf{R}$ and $\mathbf{r}$, where $\mathbf{r} = \mathbf{x}_- - \mathbf{x}_+$ is the displacement vector, and make a Galilean transformation into the barycenter rest frame.
Let us write $\mathbf{r} = [r(t), \theta(t), \phi(t)]$ in spherical coordinates.
As in the nonexotic Kepler problem, the Runge-Lenz vector is conserved, and the trajectories of $m_+$ and $m_-$ both lie in the same plane $\theta=\theta_0 = \pi/2$ without loss of generality.
The trajectory described by $r(t)$ and $\phi(t)$ obeys
\begin{align}
    E &= \frac{1}{2} \mu \dot{r}^2 + \frac{J^2}{2\mu r^2} - \frac{Gm_+m_-}{r}, \label{eq:EoMr} \\
    J & = \mu r^2 \dot{\phi} \label{eq:EoMphi},
\end{align}
where $J$ is the magnitude of the total angular momentum, $E$ is the total mechanical energy, both $E$ and $J$ are constants of the motion, and the last two terms in equation (\ref{eq:EoMr}) define the effective potential.
Upon changing the independent variable from $t$ to $\phi$, equations (\ref{eq:EoMr}) and (\ref{eq:EoMphi}) have the solution
\begin{equation}
\label{eq:rsolution}
    r(\phi) = \frac{J^2}{Gm_+m_- \mu[1+e\cos(\phi - \phi_0)]}.
\end{equation}
The ellipticity, $e$, is given by
\begin{align}
\label{eq:e}
    e = \sqrt{1+\frac{2EJ^2}{G^2m_+^2m_-^2\mu}},
\end{align}
and $\phi_0$ is a constant of integration. 

The exotic two-body problem, like its nonexotic counterpart, admits three types of motion: hyperbolic with $e > 1$, parabolic with $e = 1$, and elliptical with $e < 1$ \citep{shatskiy_kepler_2011}.
Elliptical motion is relevant especially to the restricted exotic three-body problem, which is well-posed when the primary and secondary masses execute bound orbits.
Equation (\ref{eq:e}) indicates that elliptical motion is possible for $m_+ + m_- > 0$ ($\mu < 0$), which corresponds to the case $m_+ > | m_- |$ studied in the paper.
To gain some insight into elliptical orbits involving a negative mass, we specialise to circular orbits with $e = 0$.
Circular orbits are possible, when $\mathbf{x}_+$ always lies on the line segment joining $\mathbf{R}$ and $\mathbf{x}_-$.
The gravitational repulsion of $m_-$ provides the centripetal acceleration required for $m_+$ to maintain a circular orbit. 
In contrast, for circular orbits with two positive masses, the barycenter sits between the two masses.
The angular velocity vector, $\bm{\omega}$, has magnitude 
\begin{align}
    \omega &= \frac{G^2 m_+^2 m_-^2\mu}{ J^3}
\end{align}
from (\ref{eq:EoMphi}) and (\ref{eq:rsolution}).
Elliptical motion is impossible for $m_+ + m_- < 0$.
This is because $m_+ + m_- < 0$ corresponds to $\mu > 0$, $V_{\rm eff} > 0$, $E > 0$, and $e > 1$, i.e., hyperbolic motion.

\subsection{Tertiary motion}
\label{sec:2.2}
We now add an infinitesimally light third body, denoted by $m_3$, to the exotic circular two-body problem.
In a frame centered on the barycenter and rotating with angular velocity vector $\bm{\omega}$, $m_3$ experiences a Newtonian gravitational force given by 
\begin{align}
\label{eq:F3}
     \mathbf{F}_3 = & \ -\frac{Gm_+m_3(\mathbf{X}_3 - \mathbf{X}_+)}{\abs{\mathbf{X}_3 - \mathbf{X}_+}^3} -\frac{Gm_-m_3(\mathbf{X}_3 - \mathbf{X}_-) }{\abs{\mathbf{X}_3 - \mathbf{X}_-}^3}\nonumber \\ &- m_3 \bm{\omega} \times (\bm{\omega} \times \mathbf{X}_3) - 2m_3 \bm{\omega} \times \dot{\mathbf{X}}_3,   
\end{align}
where $\mathbf{X}_+, \mathbf{X}_-$, and $ \mathbf{X}_3$ are the position vectors of $m_+, m_-$, and $m_3$, respectively, and the third and fourth terms on the right-hand side are the centrifugal and Coriolis terms, respectively.

Quantities of interest in the restricted three-body problem include the location and linear stability of stationary points of $m_3$ in the rotating frame.
Stationary points, where one has $\mathbf{F}_3 = 0$ and $\dot{\mathbf{X}}_3 = 0$, are referred to as Lagrange points \citep{lagrange_oeuvres_1873,szebehely_theory_2012,musielak_three-body_2014}.
Let us define a Cartesian coordinate system, with unit vectors $(\mathbf{e}_x, \mathbf{e}_y, \mathbf{e}_z)$, in the rotating frame centered on $\mathbf{R}$. 
Without loss of generality, we write $\bm{\omega} = \omega \mathbf{e}_z$ and place $m_+$ and $m_-$ on the $y$-axis.
Expanding (\ref{eq:F3}) out, we then find that $F_{3x}$ and $F_{3z}$ are related by
\begin{equation}
\label{eq:Lagrange_regions}
    F_{3x} X_{3z} =  F_{3z} X_{3x} + \frac{Gm_3(m_++m_-)X_{3x} X_{3z}}{\abs{\mathbf{X}_- - \mathbf{X}_+}^3}.
\end{equation}
With $F_{3x} = F_{3z} = 0$, the third term in equation (\ref{eq:Lagrange_regions}) has to be zero. 
In addition one has $m_+ + m_- \neq 0$, so every Lagrange point must satisfy $X_{3x} = 0$ or $X_{3z} = 0$. 
That is, Lagrange points are found at three locations: (a) $\mathbf{X}_3 = X_{3x} \mathbf{e}_x + X_{3y} \mathbf{e}_y$, with $X_{3x}, X_{3y} \neq 0$, (b) $\mathbf{X}_3 = X_{3y} \mathbf{e}_y + X_{3z} \mathbf{e}_z$, with $X_{3y}, X_{3z} \neq 0$, or (c) $\mathbf{X}_3 = X_{3y} \mathbf{e}_y$, with $X_{3y} \neq 0$.
Locations (a) and (c) are coplanar with $m_+$ and $m_-$, while (b) is not.

\section{Location of Lagrange points}
\label{sec:3}
\begin{figure}
    \centering
    \includegraphics[width=0.45\textwidth]{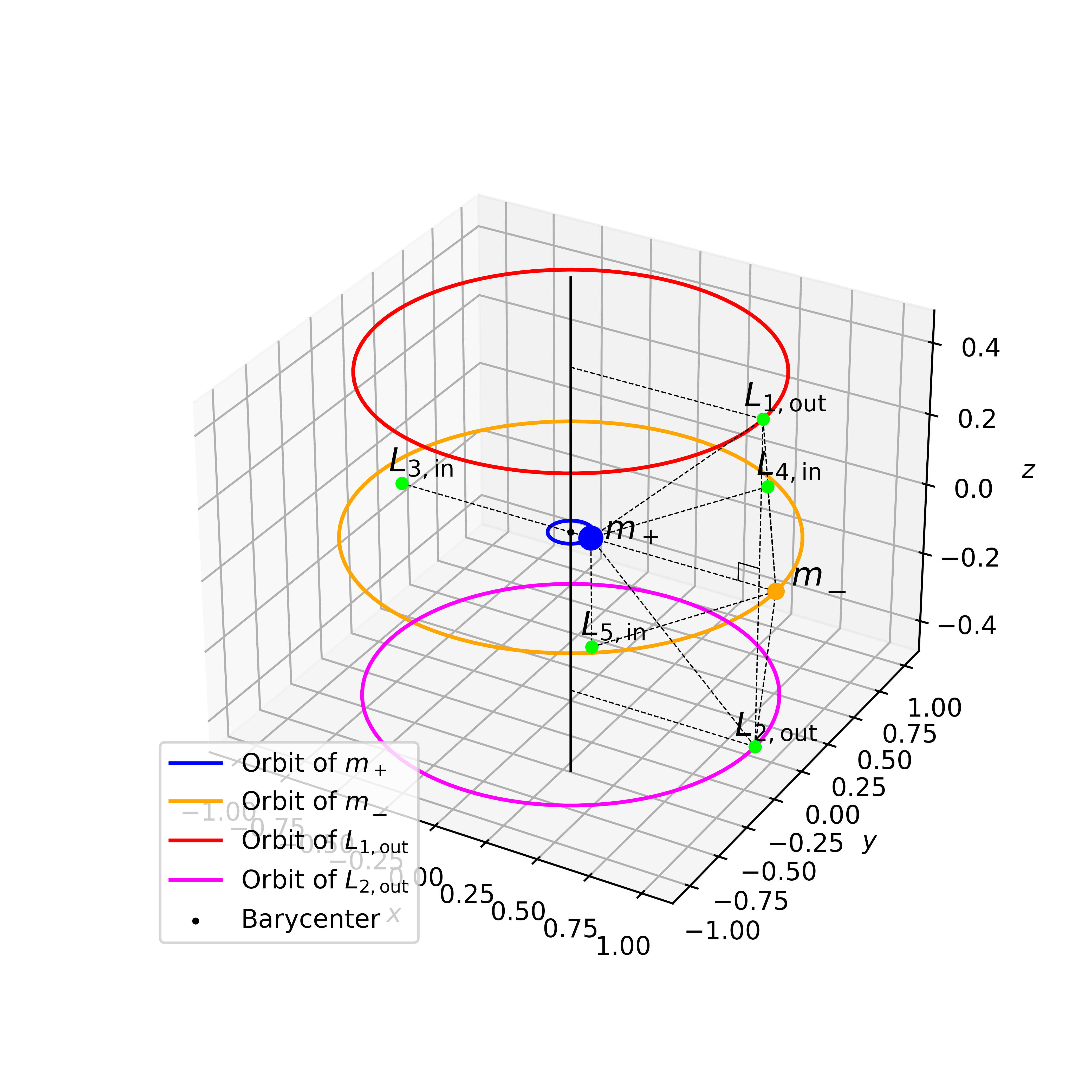}
    \caption{The Lagrange points (lime dots, labelled $L_{1, {\rm out}}, L_{2, {\rm out}}, L_{3, {\rm in}}, L_{4, {\rm in}}$ and $L_{5, {\rm in}}$) of the exotic restricted three-body problem viewed in an inertial frame. The primary and secondary masses are drawn as blue and orange dots, respectively. The black dot is the barycenter of $m_+$ and $m_-$. The blue and orange circles are the circular orbits of $m_+$ and $m_-$, respectively. The red and purple circles are the circular orbits of the out-of-plane Lagrange points, $L_{1, {\rm out}}$ and $L_{2, {\rm out}}$, respectively. Lengths are expressed in units of $\abs{\mathbf{X}_-}$. Parameter: $m_-/m_+ = -0.1$.}
    \label{fig:1}
\end{figure}

We find five solutions to $\mathbf{F}_3 = 0$ and $\dot{\mathbf{X}}_3=0$. 
Three Lagrange points are coplanar with $m_+$ and $m_-$; they are discussed in Section \ref{sec:3.1}.
Two are not; they are discussed in Section \ref{sec:3.2}. 
In the nonexotic restricted three-body problem with only positive masses, by contrast, all Lagrange points are coplanar with the primary and secondary masses.
Figure \ref{fig:1} illustrates the locations of the Lagrange points and some of their orbits in an inertial frame.
The existence of all five Lagrange points is proved in Appendix \ref{appendix:A}.

\subsection{Coplanar}
\label{sec:3.1}
The locations of the coplanar Lagrange points in Figure 1 form a one-parameter family, which is a function of the mass ratio $m_- / m_+$, when lengths are expressed in units of $\abs{\mathbf{X}_-}$.
The location of the two Lagrange points with $X_{3x} \neq 0$ and $X_{3y} \neq 0$ can be written down analytically and generally as (see Appendix \ref{appendix:A1})
\begin{equation}
\label{eq:L45}
    \mathbf{X}_3 = \pm \frac{\sqrt{3}}{2}\left(1+\frac{m_-}{m_+}\right) \mathbf{e}_x + \frac{1}{2}\left(1-\frac{m_-}{m_+}\right)\mathbf{e}_y.
\end{equation}
The points in equation (\ref{eq:L45}) separately form equilateral triangles with $m_+$ and $m_-$.
They are analogous to the two Lagrange points that form an equilateral triangle with the primary and secondary positive masses in the nonexotic restricted three-body problem, typically labelled as $L_4$ and $L_5$.
We follow a similar labelling convention here and name them $L_{4, {\rm in}}$ and $L_{5, {\rm in}}$, where the subscript, ``in'', refers to the coplanar nature (i.e.\ in the plane) of the Lagrange points.

\begin{figure}
    \centering
    \includegraphics[width=0.45\textwidth]{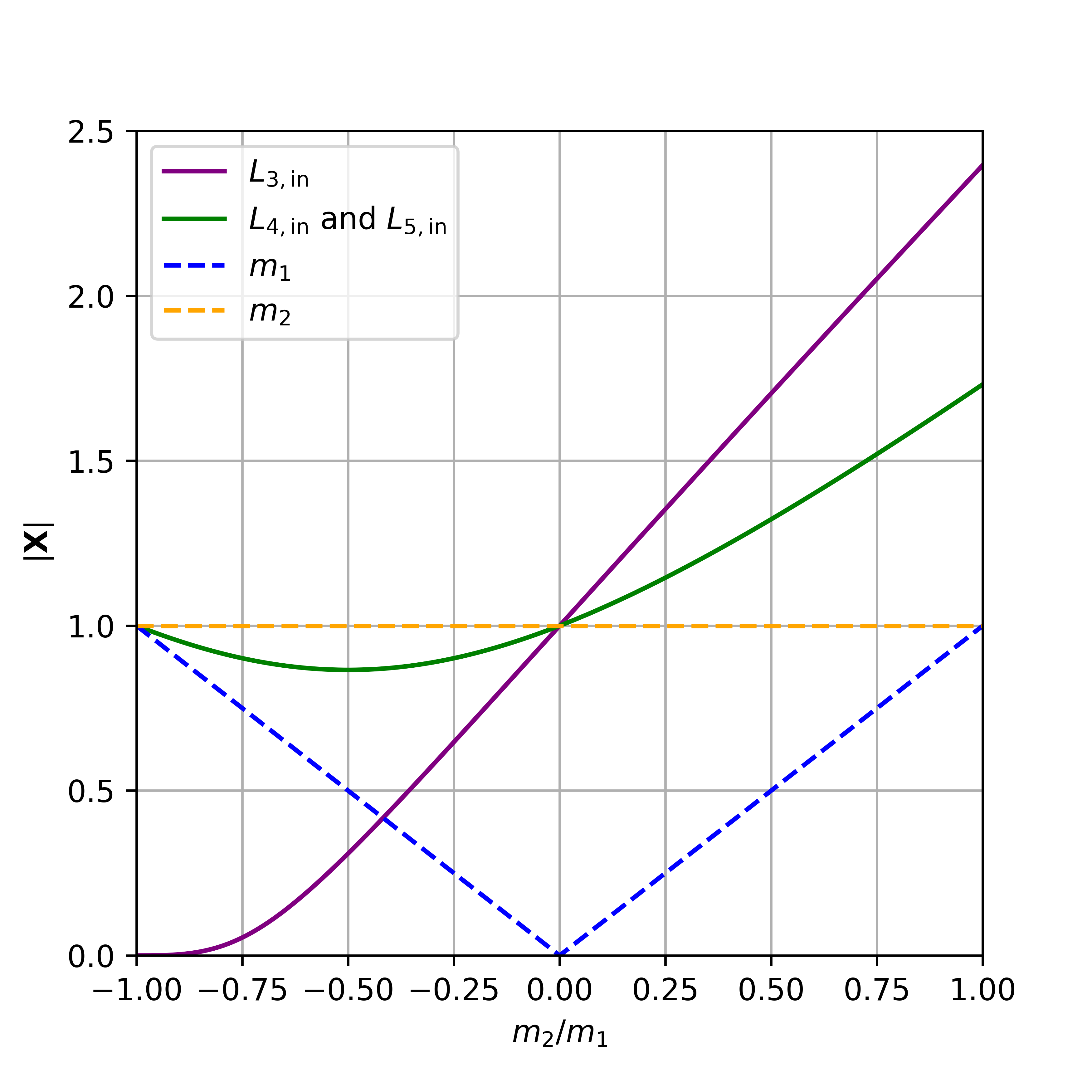}
    \caption{Distances, $\abs{\mathbf{X}}$, of the Lagrange points $L_{3, {\rm in}} ({\rm purple}),$ $L_{4, {\rm in}}$ and $L_{5, {\rm in}} ({\rm green})$, and $m_1 ({\rm dashed \ blue})$ and $m_2 ({\rm dashed \ orange})$ from the barycenter, plotted against the mass ratio $m_2/m_1$ for both exotic (left half of the plot) and nonexotic (right half of the plot) restricted three-body problems, where $m_1$ and $m_2$ are the primary and secondary masses, respectively. Lengths are dimensionless with length-scale $\abs{\mathbf{X}_2}$, where $\mathbf{X}_2$ is the position vector of $m_2$.}
    \label{fig:2}
\end{figure}

Figure \ref{fig:1} indicates that $L_{4, {\rm in}}$ and $L_{5, {\rm in}}$ are closer to the barycenter than $m_-$. 
The opposite is true in the nonexotic restricted three-body problem, where $L_{4, {\rm in}}$ and $L_{5, {\rm in}}$ are further away from the barycenter than the secondary positive mass. 
In Figure \ref{fig:2}, we plot the dimensionless distances of the Lagrange points from the barycenter, for both the exotic and nonexotic restricted three-body problems, as a function of the control parameter $m_2 / m_1$, where $m_1$ and $m_2$ are the primary and secondary masses\footnote{We introduce new symbols $m_1$ and $m_2$ for the primary and secondary masses to cover simultaneously the different signs of the exotic and nonexotic problems. In the exotic problem, we have $m_1 = m_+$ and $m_2 = m_-$.}.
Indeed, we see that $L_{4, {\rm in}}$ and $L_{5, {\rm in}}$ are closer to the barycenter for $m_2 < 0$ (left half of plot) than for $m_2 > 0$ (right half of plot).

The location of the Lagrange point with $X_{3x} = 0 = X_{3z}$ [i.e., type (c) in Section \ref{sec:2.2}] is found by solving the quintic in equation (\ref{eq:regionc2}). 
We name it $L_{3, {\rm in}}$, by analogy with the nonexotic restricted three-body problem.
The purple curve in Figure \ref{fig:2} shows the absolute value of the quintic's numerical root, $X_{3y}$, as a function of the control parameter $m_2/m_1$.
$L_{3, {\rm in}}$ is closer to the barycenter than $m_-$, as illustrated in Figures \ref{fig:1} and \ref{fig:2}.
In the nonexotic restricted three problem, by contrast, $L_{3, {\rm in}}$ is further away from the barycenter than the secondary positive mass. 
Indeed, Figure \ref{fig:2} implies $\abs{\mathbf{X}_3} < \abs{\mathbf{X}_2}$ for $m_2 < 0$ and  $\abs{\mathbf{X}_3} > \abs{\mathbf{X}_2}$ for $m_2 > 0$.
Also, the barycenter always lies on the line segment joining $L_{3, {\rm in}}$ and $m_+$ for the exotic restricted three-body problem, while the primary mass always lies on the line segment joining the barycenter and $L_{3, {\rm in}}$ in the nonexotic restricted three-body problem.
From Figure \ref{fig:2}, we see that $L_{3, {\rm in}}$ is near the barycenter for $m_+ \approx \abs{m_-}$ and near the orbit of $m_-$ for $m_+ \gg \abs{m_-}$.

In the limit where the magnitude of the primary mass greatly exceeds the magnitude of the secondary mass, $m_+ \gg \abs{m_-} \gg \abs{m_3}$, we show in Appendix \ref{appendix:A3}, that $L_{3, {\rm in}}$ is found at 
\begin{equation}
    \mathbf{X}_3 \approx - \left(1 + \frac{17}{12}\frac{m_-}{m_+}\right) \mathbf{e}_y,
\end{equation}
where $\mathbf{X}_3$ is expressed in units of $\abs{\mathbf{X}_2}$.


\subsection{Out-of-plane}
\label{sec:3.2}

\begin{figure}
    \centering
    \includegraphics[width=0.45\textwidth]{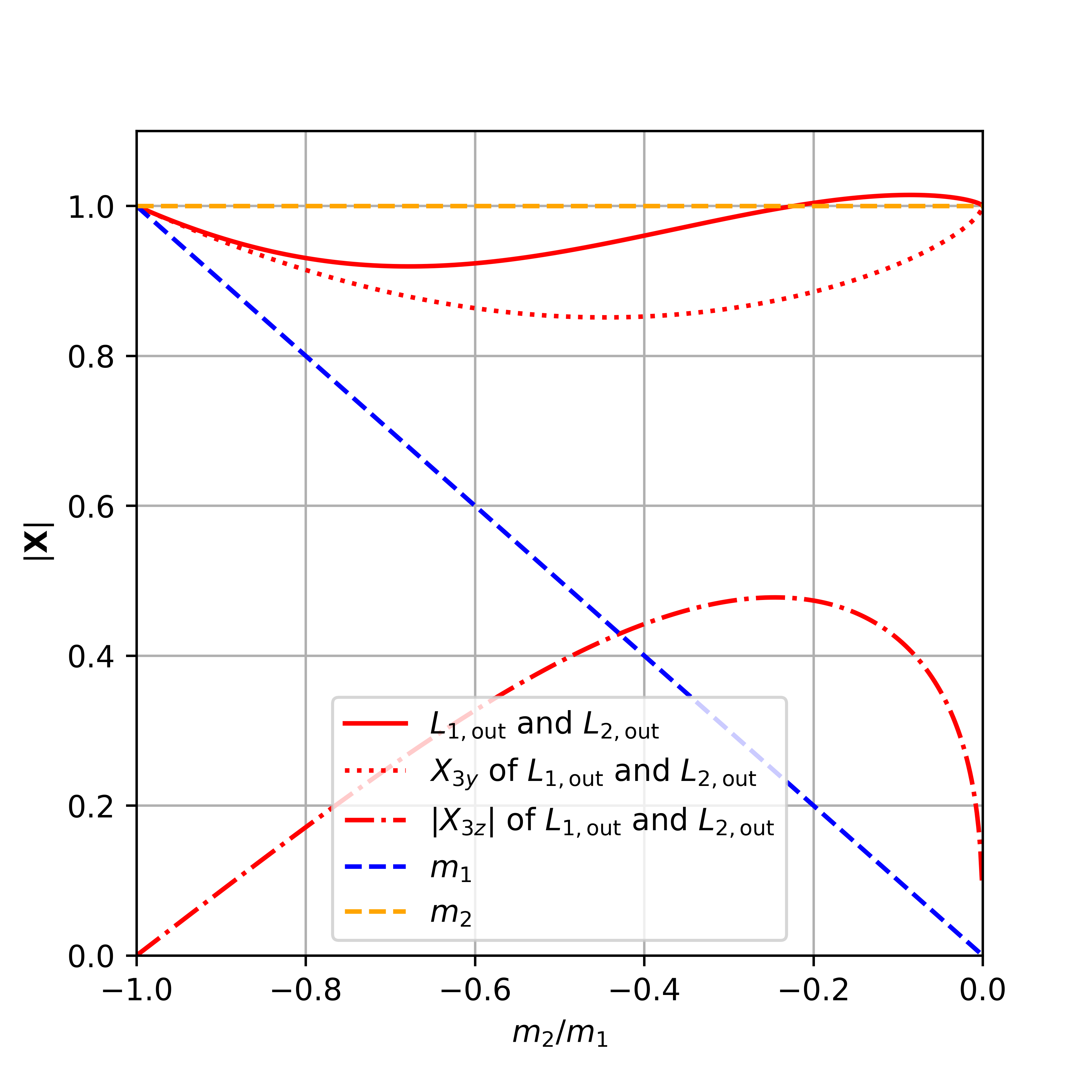}
    \caption{As in Figure \ref{fig:2}, but for the out-of-plane Lagrange points $L_{1, {\rm out}}$ and $L_{2, {\rm out}}$ (solid and broken red curves) of the exotic restricted three-body problem. There are no analogous Lagrange points for the nonexotic restricted three-body problem. $X_{3y}$ and $\abs{X_{3z}}$ of $L_{1, {\rm out}}$ and $L_{2, {\rm out}}$ are plotted as dotted and dash-dotted red curves, respectively.}
    \label{fig:3}
\end{figure}
We show in Appendix \ref{appendix:A2}, that two Lagrange points are located at $X_{3y},\, X_{3z} \neq 0$ [i.e., type (b) in Section \ref{sec:2.2}], at the same $X_{3y}$ and opposite $X_{3z}$. 
These Lagrange points lie above and below the plane containing the orbits of $m_+$ and $m_-$.
At these points, the gravitational repulsion of $m_-$ out of the plane balances the gravitational attraction of $m_+$ into the plane.
No analogous Lagrange points exist in the nonexotic restricted three-body problem. 
However, for notational simplicity, we name them $L_{1, {\rm out}}$ and $L_{2, {\rm out}}$. 

In Figure \ref{fig:3}, we plot the dimensionless distance from the barycenter, $\abs{\mathbf{X}}$, of $L_{1, {\rm out}}$ and $L_{2, {\rm out}}$ (which are equal; solid red curve), $X_{3x}$ (dotted red curve), $X_{3z}$ (dash-dotted red curve), $m_1$ (blue dashed curve) and $m_2$ (orange dashed curve) as functions of the control parameter $m_2/m_1$.
We only plot $-1 < m_2/m_1 \leq 0$ because no analogy exists for the nonexotic restricted three-body problem.
$L_{1, {\rm out}}$ and $L_{2, {\rm out}}$ in Figure \ref{fig:3} are obtained by numerically solving the quintic (\ref{eq:regionb}).
Figure \ref{fig:3} shows that $L_{1, {\rm out}}$ and $L_{2, {\rm out}}$ have $0.91 \lesssim \abs{\mathbf{X}_3}/\abs{\mathbf{X}_-} \lesssim 1.02 $; that is, these two Lagrange points are approximately as far apart as $m_-$ from the barycenter.
Unlike $L_{3, {\rm in}}$, $L_{4, {\rm in}}$ and $L_{5, {\rm in}}$, $L_{1, {\rm out}}$ and $L_{2, {\rm out}}$ can be further away than $m_-$ from the barycenter for some $m_-/m_+$, as shown in Figure \ref{fig:3}; we find $\abs{\mathbf{X}_3} \gtrsim \abs{\mathbf{X}_-}$ for $-0.22 \lesssim m_-/m_+ < 0$ for $L_{1, {\rm out}}$ and $L_{2, {\rm out}}$.
We also find $\abs{X_{3z}} \lesssim 0.48$.

In the regime $m_+ \gg \abs{m_-} \gg \abs{m_3}$, we show in Appendix \ref{appendix:A3}, that $L_{1, {\rm out}}$ and $L_{2, {\rm out}}$ are found at
\begin{align}
    \mathbf{X}_3 \approx & \left[1 - \frac{3}{8}\left(-\frac{m_-}{m_+}\right)^{2/3}\right] \mathbf{e}_y \nonumber \\ &\pm \left[\left(-\frac{m_-}{m_+}\right)^{1/3} + \frac{7}{128} \frac{m_-}{m_+}\right] \mathbf{e}_z,
\end{align}
where $\mathbf{X}_3$ is expressed in units of $\abs{\mathbf{X}_2}$.

\section{Linear stability of Lagrange points}
\label{sec:4}
We now study the linear stability of $m_3$ when perturbed away from the five Lagrange points.  
In Section \ref{sec:4.1}, we linearize the equations of motion of $m_3$ in the vicinity of the Lagrange points.
We analyze the eigenvalues and discuss stability in terms of an effective potential in Section \ref{sec:4.2}.

\subsection{Linearized equations of motion}
\label{sec:4.1}
Suppose that $m_3$ is perturbed slightly from a Lagrange point located at $\mathbf{L}_k$, with $1 \leq k \leq 5$.
We write
\begin{equation}
    \mathbf{X}_3 = \mathbf{L}_k + \delta\mathbf{X}
\end{equation}
and
\begin{equation}
    \mathbf{V}_3 = \delta \dot{\mathbf{X}},
\end{equation}
where $\mathbf{V}_3 = \dot{\mathbf{X}}_3$ is the velocity of $m_3$ in the rotating frame.
Combining $\mathbf{X}_3$ and $\mathbf{V}_3$ into 
\begin{equation}
    \mathbf{A} = \begin{bmatrix}
        \mathbf{X}_3 \\
        \mathbf{V}_3
    \end{bmatrix}
\end{equation}
and Taylor expanding about $\mathbf{L}_k$, we get
\begin{equation}
    \frac{dA_a}{dt} = U_{ab} A_b,
\end{equation}
with $1 \leq a, b \leq 6$, where $\mathbf{U}$ is a $6 \times 6$ matrix given by
\begin{equation}
    \mathbf{U} = \begin{bmatrix}
        \mathbf{0} & \mathbf{I} \\
        \mathbf{B} & \mathbf{C} 
    \end{bmatrix},
\end{equation}
and $\mathbf{0}$ and $\mathbf{I}$ are $3 \times 3$ zero and identity matrices, respectively.
The Cartesian components of $\mathbf{B}$ and $\mathbf{C}$ are given by
\begin{equation}
    B_{ij} = \frac{1}{m_3} \frac{\partial F_{3i}}{\partial X_{3j}},
\end{equation}
and
\begin{equation}
    C_{ij} = \frac{1}{m_3} \frac{\partial F_{3i}}{\partial V_{3j}},
\end{equation}
with $i, j \in \{x, y, z\}$.

\subsection{Eigenvalues}
\label{sec:4.2}

\begin{figure}
    \begin{subfigure}[b]{\columnwidth}
    	\includegraphics[width=\textwidth]{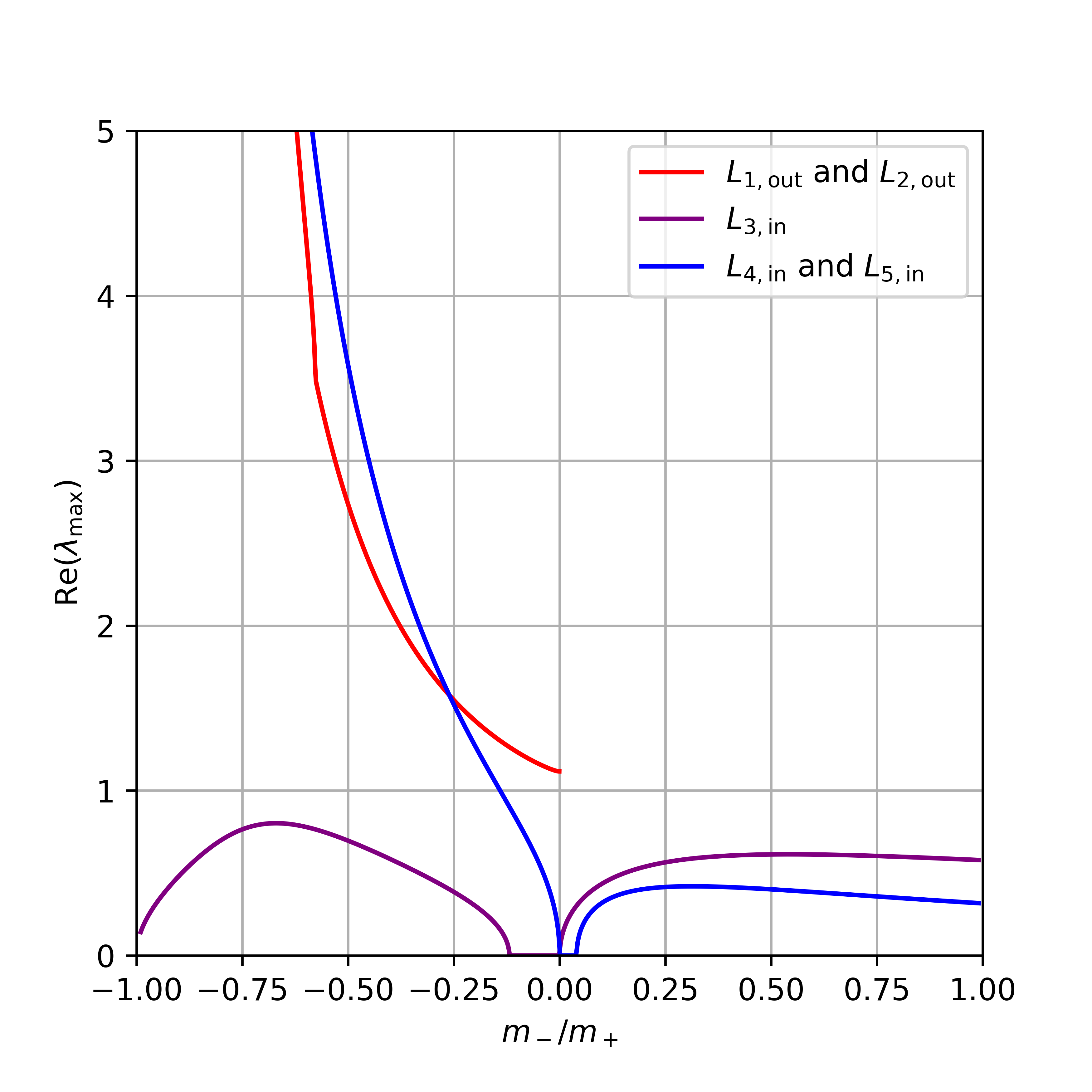}
    	\caption{}
    	\label{fig:4a}
    \end{subfigure}
    \hfill
    \begin{subfigure}[b]{\columnwidth}
    	\includegraphics[width=\textwidth]{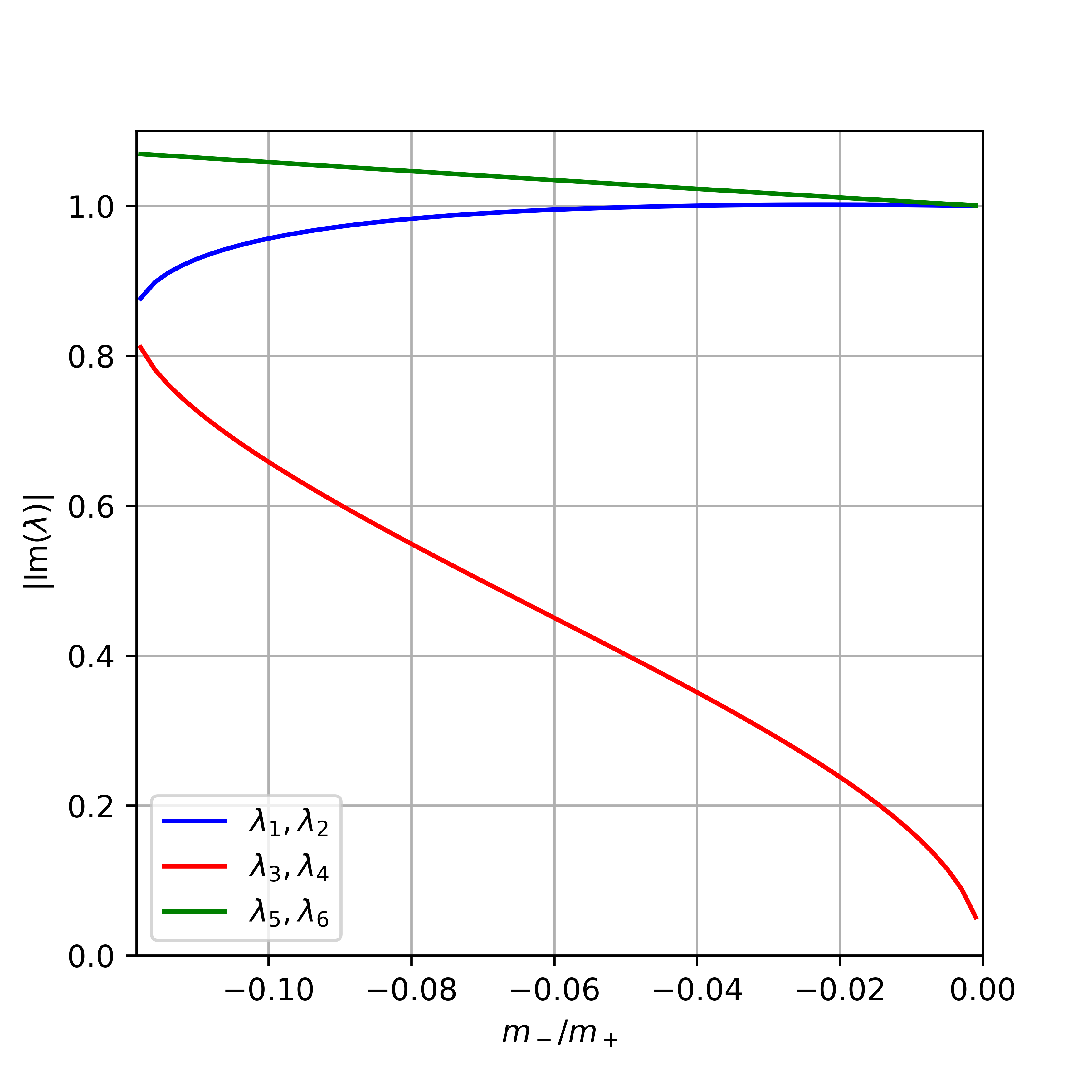}
    	\caption{}
    	\label{fig:4b}
    \end{subfigure}
    \hfill
    \caption{Eigenvalues, $\lambda$, of $\mathbf{U}/(Gm_+)$ versus $m_-/m_+$. (a) Largest positive real eigenvalue, ${\rm Re}(\lambda_{\rm max})$, at the Lagrange points. The colour of the curves in Figure \ref{fig:5a} is the same as in Figures \ref{fig:2} and \ref{fig:3}. (b) Absolute value of the imaginary component of the eigenvalues of $L_{3, {\rm in}}$, ${\rm Im}(\lambda)$. The blue, red and green curves in Figure \ref{fig:5b} represent the three conjugate eigenvalue pairs. Lengths and time are expressed in units of $\abs{\mathbf{X}_-}$ and $\omega^{-1}$, respectively, and $\lambda$ has units of $(Gm_+)^{1/2}$. The kink in the red curve ($L_{1, {\rm out}}$ and $L_{2, {\rm out}}$) in Figure \ref{fig:4a} is where four of the six eigenvalues' real components are equal.}
    \label{fig:4}
\end{figure}

For arbitrary $m_-/m_+$, one can calculate the eigenvalues of $\mathbf{U}/(Gm_+)$ for $L_{4, {\rm in}}$ and $L_{5,{\rm in}}$ analytically but must calculate the eigenvalues of $L_{1, {\rm out}}, \,  L_{2, {\rm out}}, \, L_{3, {\rm in}}$ numerically. For example, the transition of $m_3$ in $L_{4, {\rm in}}$ and $L_{5, {\rm in}}$ from instability to linear stability  in Figure \ref{fig:4a} occurs when $0< m_-/(m_-+m_+) < 1/2[1-(23/27)^{1/2}]$.
Upon doing so, we find that $L_{1, {\rm out}}$, $L_{2, {\rm out}}$, $L_{4, {\rm in}}$ and $L_{5, {\rm in}}$ have at least one eigenvalue that is real and positive.
Hence $L_{1, {\rm out}}$, $L_{2, {\rm out}}$, $L_{4, {\rm in}}$ and $L_{5, {\rm in}}$ are linearly unstable.
On the other hand, $L_{3, {\rm in}}$ has strictly imaginary eigenvalues, i.e.\ three complex conjugate pairs for $m_+ \gtrsim 8.4 \abs{m_-}$, and hence is linearly stable.
The largest positive real eigenvalue, ${\rm Re}(\lambda_{\rm max})$, is plotted versus $m_- / m_+$ in the top panel of Figure \ref{fig:4} for the Lagrange points.
We see from Figure \ref{fig:4a} that $L_{4, {\rm in}}$ and $L_{5, {\rm in}}$ are linearly unstable for the exotic restricted three-body problem but can be linearly stable for the nonexotic restricted three-body problem.
The time-scale for drifting away from the Lagrange points is $(1+m_-/m_+)[(Gm_+)^{1/2} {\rm Re}(\lambda_{\rm max})]^{-1}$, where ${\rm Re}(\lambda_{\rm max})$ ranges from $1/5$ to $\infty \, (Gm_+)^{1/2}$ across the plotted range, and time and length are expressed in $\omega^{-1}$ and $\abs{\mathbf{X_-}}$, respectively. 
The magnitudes of the three imaginary eigenvalue pairs, ${\rm Im}(\lambda)$, for $L_{3,{\rm in}}$ are plotted versus $m_+/m_-$ in the bottom panel of Figure \ref{fig:4} for $m_+ \gtrsim 8.4 \abs{m_-}$. 
The periods of the three independent oscillations executed by $m_3$ in the vicinity of $L_{3,{\rm in}}$ are $(1+m_-/m_+)[(Gm_+)^{1/2} {\rm Im
}(\lambda_{\rm max})]^{-1}$, where ${\rm Im}(\lambda)$ ranges from $0.94$ to $20\,(Gm_+)^{1/2}$  across the plotted range. 
The locations of the Lagrange points, except for $L_{4, {\rm in}}$ and $L_{5, {\rm in}}$, are computed numerically from Section \ref{sec:3}. 

\begin{figure*}
    \begin{subfigure}{\columnwidth}
    	\includegraphics[width=0.85\textwidth]{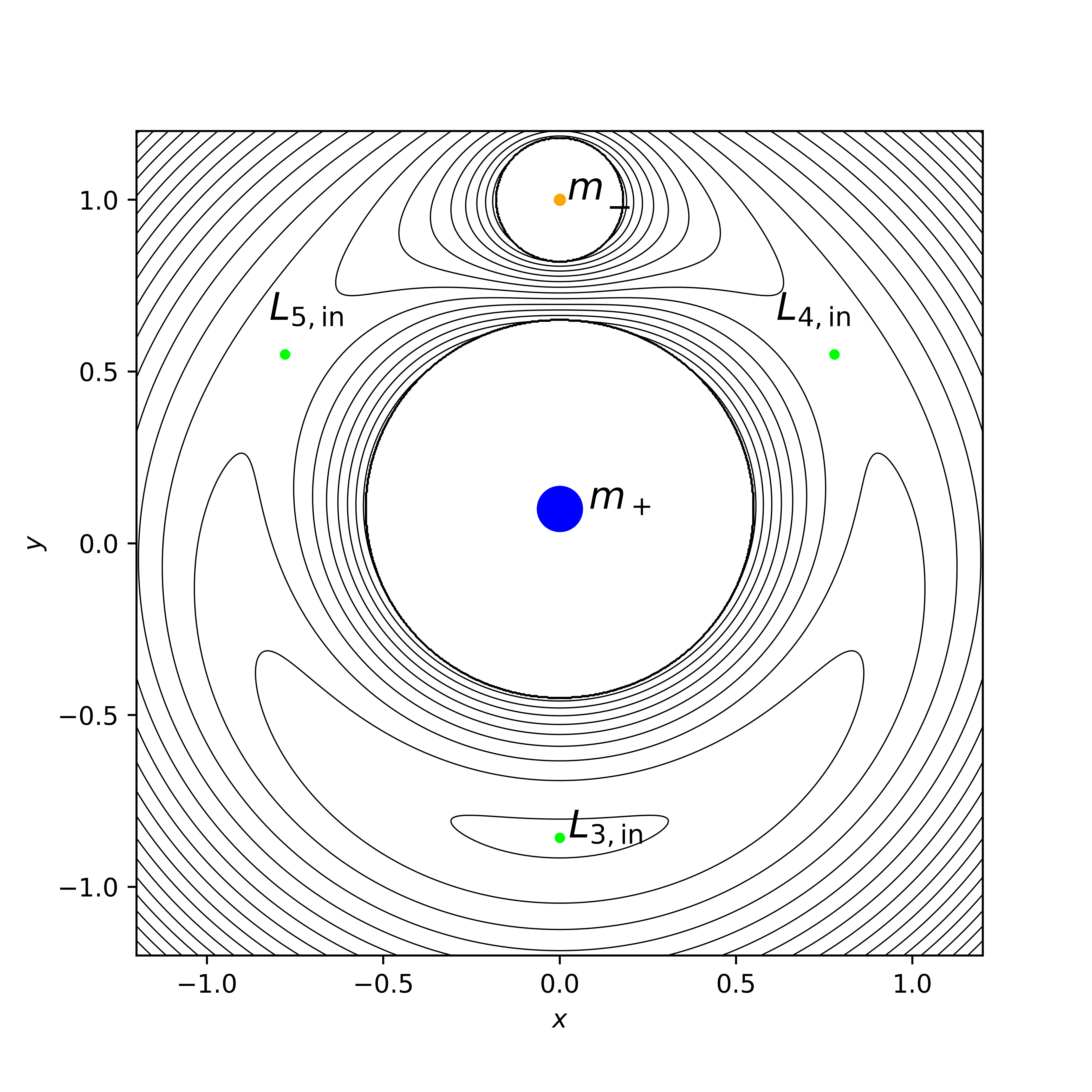}
    	\caption{}
    	\label{fig:5a}
    \end{subfigure}
    \begin{subfigure}[b]{\columnwidth}
    	\includegraphics[width=0.85\textwidth]{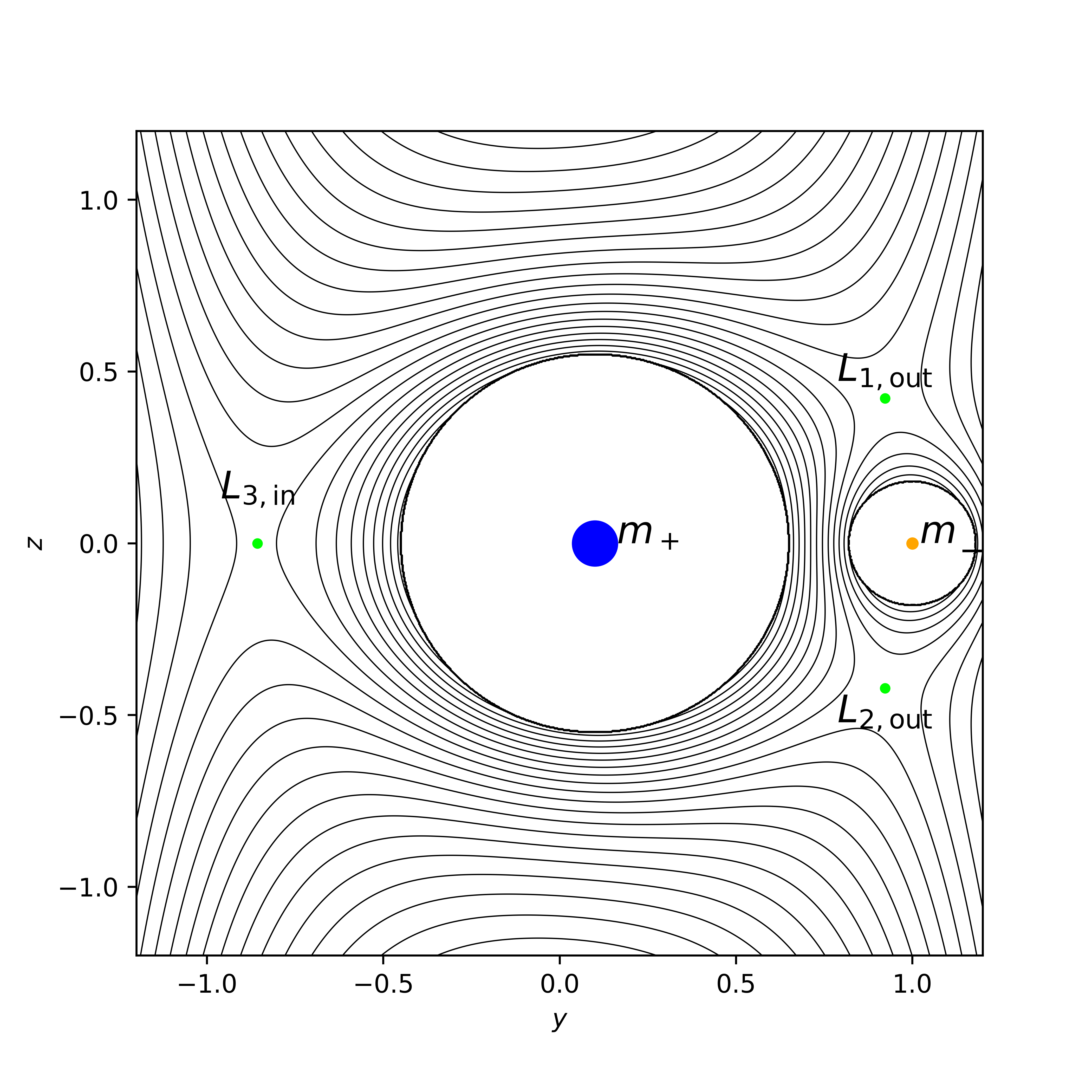}
    	\caption{}
    	\label{fig:5b}
    \end{subfigure}
    \hfill
    \vfill
    \begin{subfigure}[b]{\columnwidth}
    	\includegraphics[width=0.85\textwidth]{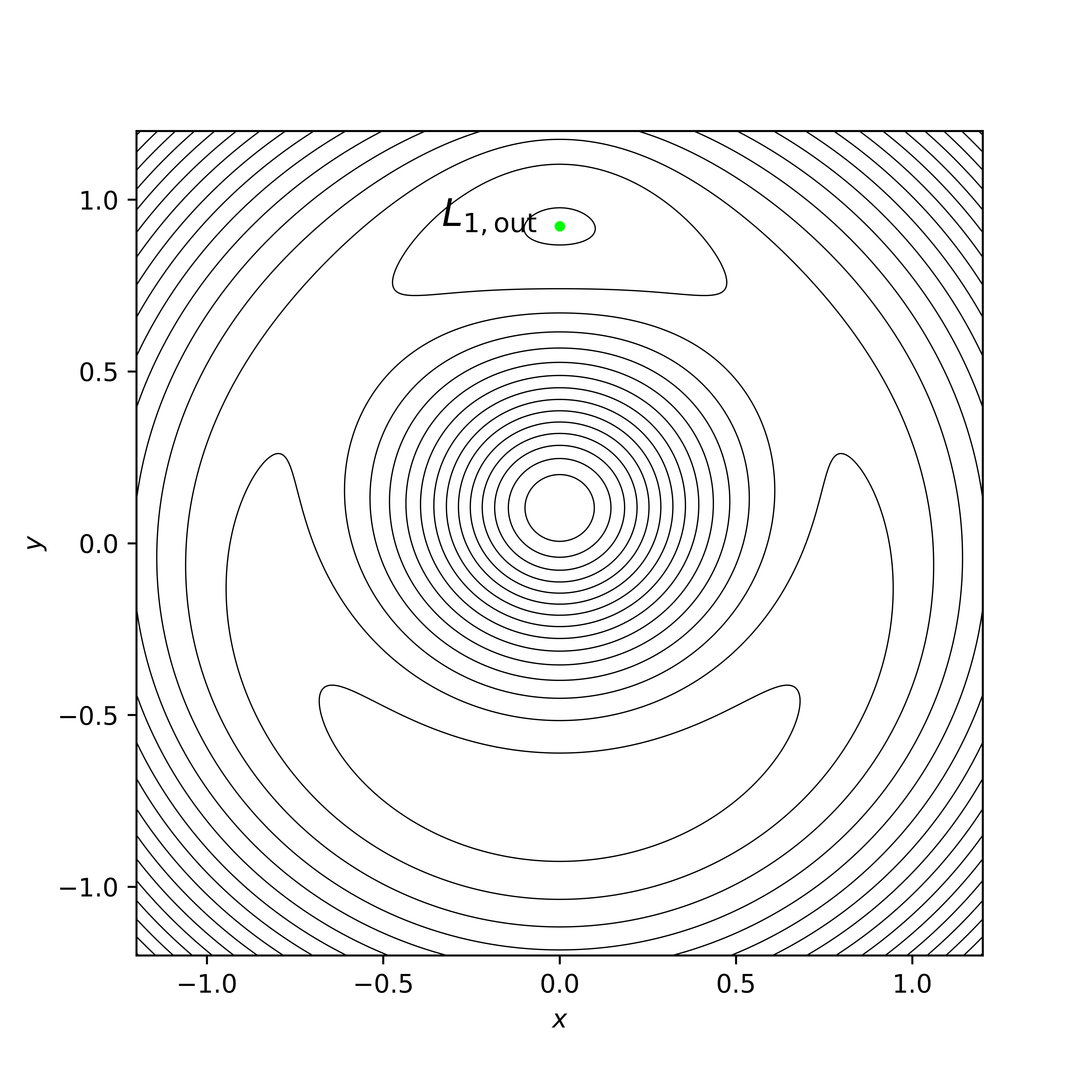}
            \centering
    	\caption{}
    	\label{fig:5c}
    \end{subfigure}
    \hfill
    \caption{Contour plots of cross-sections of the effective potential, $V_{\rm eff}$, of the exotic restricted three-body problem in the frame rotating with angular velocity $\bm{\omega}$ centered on the barycenter. (a) Cross-section in the plane normal to ${\mathbf n} = {\mathbf e}_z$ at $z = 0$. (b) Cross-section in the plane normal to ${\mathbf n} = {\mathbf e}_x$ at $x = 0$. (c) Cross-section in the plane normal to ${\mathbf n} = {\mathbf e}_z$ at $z = X_{3z}$ for $L_{1, {\rm out}}$. Solid curves are for where $V_{\rm eff}$ is positive. Circles around $m_+$ and $m_-$, including where $V_{\rm eff}$ is positive near $m_-$, are cut out to aid visualisation of $V_{\rm eff}$ around the Lagrange points. Lengths are expressed in units of $\abs{\mathbf{X}_-}$. Parameter: $m_-/m_+ = -0.1$.}
    \label{fig:5}
\end{figure*}

To assist with visualizing the stability properties of the Lagrange points, we plot the effective potential $V_{\rm eff}$ in Figure 5. 
Specifically, in the frame rotating with $\bm \omega$, we have
\begin{align}
\label{eq:Veff}
    \frac{V_{\rm eff}}{Gm_+m_3} =& -\frac{1}{\abs{\mathbf{X}_3 + m_-/m_+}}-\frac{m_-/m_+}{\abs{\mathbf{X}_3 - 1}} \nonumber \\ &- \frac{ X_{3x}^2 + X_{3y}^2}{2 (1 + m_-/m_+)^2}
\end{align}
The gradient of equation (\ref{eq:Veff}) reproduces the fictitious centrifugal force but not the Coriolis force. 
Contours of cross-sections of $V_{\rm eff}$ for $m_-/m_+ = -0.1$ are plotted in the planes with unit normals $\mathbf{n} = \mathbf{e}_z$ at $z = 0$ and $\mathbf{n} = \mathbf{e}_x$ at $x=0$, $\mathbf{n} = \mathbf{e}_z$ at $z = X_{3z}$ of $L_{1,{\rm out}}$ in Figures \ref{fig:5a}, \ref{fig:5b}, and \ref{fig:5c}, respectively. 
The contours of $V_{\rm eff}$ with $\mathbf{n} = \mathbf{e}_z$ at $z = X_{3z}$ of $L_{2,{\rm out}}$ equal those of $L_{1,{\rm out}}$.
We see that $L_{1,{\rm out}}$ and $L_{2,{\rm out}}$ are located on saddle points in Figure \ref{fig:5b} and on a ``hilltop'' in Figure \ref{fig:5c}, consistent with being unstable.
Likewise, $L_{4,{\rm in}}$ and $L_{5,{\rm in}}$ are positioned on saddle points in Figure \ref{fig:5a}.
Finally, $L_{3, {\rm in}}$ is positioned on a hill in Figure \ref{fig:5a} and on a saddle point in Figure \ref{fig:5b}.
However, it is stable for $m_+ \gtrsim 8.4 \abs{m_-}$, when one includes the velocity-dependent Coriolis force\footnote{Analogously, $L_4$ and $L_5$ of the nonexotic restricted three-body problem can be linearly stable despite sitting on the hill of the effective potential.}.

\section{Conclusion}
\label{sec:5}
In this paper, we study the Newtonian restricted three-body gravitational problem with a positive primary and a negative secondary point mass.
The negative point masses in this paper are assumed to have the same gravitational and inertial mass, thereby satisfying the principle of equivalence.
We find five Lagrange points in Section \ref{sec:3}, three of which are coplanar with $m_+$ and $m_-$, while two are not. 
The three coplanar Lagrange points have counterparts in the nonexotic restricted three-body problem, where two points form separate equilateral triangles with $m_+$ and $m_-$ and one is collinear with $m_+$ and $m_-$. 
In comparison, the two out-of-plane points, $L_{1, {\rm out}}$ and $L_{2, {\rm out}}$, do not have nonexotic counterparts.
$L_{1, {\rm out}}$ and $L_{2, {\rm out}}$ exist because $m_-$ repels $m_3$.
They are found at a distance $0.91 \lesssim \abs{\mathbf{X}_3} \lesssim 1.02$ from the barycenter, where distances are expressed in the units of $\abs{\mathbf{X}_-}$.
The Lagrange point coplanar and collinear with $m_+$ and $m_-$, $L_{3, {\rm in}}$, is linearly stable for $m_+ \gtrsim 8.4 \abs{m_-}$ and unstable otherwise.
The other four Lagrange points are unstable for all $m_-/m_+$.

The results in Sections \ref{sec:3} and \ref{sec:4} showcase the following counterintuitive properties of gravitationally repulsive matter.
(i) $m_+$ and $m_-$ can form elliptical orbits for $m_+ > \abs{m_-}$ [see Section \ref{sec:2.1} and \citet{shatskiy_kepler_2011}], such that $m_+$ lies between the barycenter and $m_-$. 
(ii) $L_{1, {\rm out}}$ and $L_{2, {\rm out}}$ are not coplanar with $m_+$ and $m_-$; the plane of the orbit of the tertiary mass is parallel to, yet above or below, the plane containing the orbits of the primary and secondary masses.
(iii) All coplanar Lagrange points are closer to the barycenter than $m_-$, unlike in the nonexotic problem, where four out of five coplanar Lagrange points are further away from the barycenter than the secondary mass.
(v) The out-of-plane Lagrange points can be further away from the barycenter than $m_-$ for $-0.22 \lesssim m_-/m_+ < 0$ (see Figure \ref{fig:3}).
(vi) $L_{3, {\rm in}}$ is linearly stable for all $m_+ \gtrsim 8.4\abs{m_-}$, unlike in the nonexotic problem.
Point (ii) is perhaps the most intriguing, as $L_{1, {\rm out}}$ and $L_{2, {\rm out}}$ are in a plane containing no other masses, an unusual observational signature.

We emphasize in closing that there is no experimental evidence at the time of writing that bodies with negative mass exist. 
Nevertheless, it is useful to develop a detailed theoretical understanding of their gravitational dynamics, in preparation for a hypothetical day in the future when the experimental situation may arguably change. 
In this paper, we contribute by generalizing pioneering studies of the exotic two-body problem \citep{bondi_negative_1957,bonnor_exact_1964,shatskiy_kepler_2011} to restricted three bodies, complementing studies of three or more bodies by other authors \citep{roberts_continuum_1999,celli_homographic_2005,celli_central_2007, manfredi_cosmological_2018,farnes_unifying_2018,rahman_existence_2019}. Future work along related lines may include analyzing the gravitational interactions of extended, continuous, negative mass distributions, both in their own right and as toy classical models for bounded fields with negative energy density, which are postulated to act as sources of stress-energy in exotic spacetimes, such as traversable wormholes and warp drives \citep{morris_wormholes_1988,alcubierre_warp_1994,lobo_exotic_2007}.

\begin{acknowledgments}
Parts of this research are supported by an Australian Government Research Training Program Scholarship (Stipend), Research Training Program Scholarship (Fee Offset), Rowden White Scholarship, McKellar Prize in Theoretical Physics, Professor Kernot Research Scholarship in Physics and the Australian Research Council (ARC) Centre of Excellence for Gravitational Wave Discovery (OzGrav) (grant number CE170100004).
\end{acknowledgments}

\appendix

\section{Analysis of Lagrange point locations}
\label{appendix:A}
In this appendix, we summarize the main calculations concerning the location of the Lagrange points, whose results are presented in Section \ref{sec:3}. 
We prove the existence of the coplanar and out-of-plane Lagrange points in Appendices \ref{appendix:A1} and \ref{appendix:A2} respectively. We develop approximate formulas for the locations of $L_{1, {\rm out}}, L_{2, {\rm out}}$ and $L_{3, {\rm in}}$ in the regime $m_+ \gg |m_-| \gg m_3$ in Appendix \ref{appendix:A3}.
We follow Section \ref{sec:2} and use the Cartesian coordinate system, with unit vectors $(\mathbf{e}_x, \mathbf{e}_y, \mathbf{e}_z)$, centered on the barycenter and in a frame rotating with angular velocity $\bm{\omega}$.
For simplicity, we express lengths in units of $\abs{\mathbf{X}_-}$, such that $X_{-y} = 1$ and $X_{+y} = -m_-/m_+$.

\subsection{Existence of the coplanar points \texorpdfstring{$L_{3, {\rm in}}$}{L3in}, \texorpdfstring{$L_{4, {\rm in}}$}{L4in} and \texorpdfstring{$L_{5, {\rm in}}$}{L5in}}
\label{appendix:A1}
The three coplanar Lagrange points are found in regions (a) and (c), as specified in Section \ref{sec:2.2}.
In region (a), we have $X_{3x} \neq 0, X_{3y} \neq 0$ and $X_{3z} = 0$.
With these conditions, the $x$- and $y$-components of equation (\ref{eq:F3}) combine and simplify to give

\begin{equation}
\label{eq:A1}
    \left(X_{3y}+\frac{m_-}{m_+}\right)^2 = (X_{3y}-1)^2.
\end{equation}
Solving equation (\ref{eq:A1}) yields
\begin{equation}
\label{eq:A2}
    X_{3y} = \frac{1}{2}\left(1-\frac{m_-}{m_+}\right)
\end{equation}
Substituting equation (\ref{eq:A2}) into the $x$-component of equation (\ref{eq:F3}) yields
\begin{equation}
\label{eq:A3}
    X_{3x} = \pm \frac{\sqrt{3}}{2}\left(1+\frac{m_-}{m_+}\right).
\end{equation}
We label the Lagrange points defined by equation (\ref{eq:A3}) as $L_{4, {\rm in}}$ and $L_{5, {\rm in}}$.

In region (c), with $X_{3x} = X_{3z} = 0$ and $X_{3y} \neq 0$, the condition $F_{3y} = 0$ implies 
\begin{widetext}
\begin{align}
\label{eq:regionc}
    0 =& - (X_{3y} - 1)^2{\rm sgn}\left(X_{3y}+\frac{m_-}{m_+}\right) -\frac{m_-}{m_+} \left(X_{3y}+\frac{m_-}{m_+}\right)^2{\rm sgn}(X_{3y} - 1)  + \frac{ X_{3y} (X_{3y} - 1)^2}{(1+m_-/m_+)^{2}} \left(X_{3y}+\frac{m_-}{m_+}\right)^2.
\end{align}
\end{widetext}

For $X_{3y} > 1$, and writing $\mathcal{X}_{3y} = X_{3y} - 1$, we rearrange and obtain
\begin{widetext}
\begin{align}
\label{eq:regionc1}
    0 =& \ \mathcal{X}_{3y}^5 \left(\frac{1}{1+m_-/m_+}\right)^2 + \mathcal{X}_{3y}^4 \left[\frac{3+2m_-/m_+}{(1+m_-/m_+)^2} \right] + \mathcal{X}_{3y}^3 \left(\frac{3+m_-/m_+}{1+m_-/m_+} \right) \nonumber \\  &  + \mathcal{X}_{3y}^2 \left(-\frac{m_-}{m_+} \right) + \mathcal{X}_{3y} \left[-\frac{2m_-}{m_+}\left(1+\frac{m_-}{m_+}\right)\right] - \frac{m_-}{m_+} \left(1+\frac{m_-}{m_+}\right)^2.
\end{align}
\end{widetext}
All of the coefficients of the polynomial in $\mathcal{X}_{3y}$ are positive. 
Hence Descartes' rule of signs states that equation (\ref{eq:regionc1}) has no real positive root. 
That is, no Lagrange point exists for $X_{3y} > 1$.

Let us now consider the case $X_{3y} < -m_-/m_+$ with the substitution $\mathcal{X}_{3y} = -(X_{3y} + m_-/m_+)$.
Equation (\ref{eq:regionc}) implies 
\begin{widetext}
\begin{align}
\label{eq:regionc2}
    0 = \ &\mathcal{X}_{3y}^5 \left(-\frac{1}{1+m_-/m_+}\right)^2 + \mathcal{X}_{3y}^4 \left[-\frac{3+2m_-/m_+}{(1+m_-/m_+)^2} \right]  + \mathcal{X}_{3y}^3 \left(\frac{2}{1+m_-/m_+} -3 \right) + \mathcal{X}_{3y}^2 \nonumber \\&+ \mathcal{X}_{3y} \left[2\left(1+\frac{m_-}{m_+}\right)\right] +\left(1+\frac{m_-}{m_+}\right)^2.
\end{align}
\end{widetext}
The coefficients of $\mathcal{X}_{3y}^5, \mathcal{X}_{3y}^4$ and $\mathcal{X}_{3y}^2, \mathcal{X}_{3y}^1, \mathcal{X}_{3y}^0$ are negative and positive, respectively, while the coefficient of $\mathcal{X}_{3y}^3$ is negative for $m_-/m_+ > -1/3$ and positive for $m_-/m_+ < -1/3$. That is, there is one change of sign. 
By Descartes' rule of signs, equation (\ref{eq:regionc2}) has one real positive root. 
The root is broadly analogous to $L_3$ in the nonexotic problem, so we label it $L_{3, {\rm in}}$.

Finally, let us consider the case $-m_-/m_+ < X_{3y} < 1$. 
Equation (\ref{eq:regionc}) implies
\begin{equation}
\label{eq:A7}
    0 = -\frac{1}{(X_{3y} +m_-/m_+)^2} + \frac{m_-/m_+}{(X_{3y} - 1)^2} + \frac{X_{3y}}{(1+m_-/m_+)^{2}}.
\end{equation}
Applying the inequalities $-(X_{3y} + m_-/m_+)^{-2} < -(1+m_-/m_+)^{-2}$, $(X_{3y}-1)^{-2} < (1+m_-/m_+)^{-2}$, and $X_{3y}<1$ to the first, second, and third terms respectively, we conclude that the right-hand side of equation (\ref{eq:A7}) is negative definite. Hence no real roots exist for $-m_-/m_+ < X_{3y} < 1$.

\subsection{Existence of out-of-plane points \texorpdfstring{$L_{1, {\rm out}}$}{L1out} and \texorpdfstring{$L_{2, {\rm out}}$}{L2out}}
\label{appendix:A2}
Two out-of-plane Lagrange points are found in region (b), as specified in Section \ref{sec:2.2}. In region (b), we have $X_{3x} = 0$ and $X_{3y}, X_{3z} \neq 0$. With these conditions, the $y$- and $z$-components of equation (\ref{eq:F3}) combine and simplify to give

\begin{align}
\label{eq:regionb}
    0 =& \ \chi_{3y}^{5} 2\left(1 + \frac{m_-}{m_+} \right) + \chi_{3y}^{2} \left(-1 + \frac{m_-^2}{m_+^2} \right) \nonumber \\&+ \left(-1 + \frac{m_-^{2/3}}{m_+^{2/3}} \right) \left(1 + \frac{m_-}{m_+} \right)^2,
\end{align}
with $\chi_{3y} = X_{3y}^{1/3}$. 
The coefficient of $\chi_{3y}^5$ is positive while the coefficients of $\chi_{3y}^2$ and $\chi_{3y}^0$ are negative. 
That is, there is one change of sign for the coefficients of non-zero polynomial terms, ordered by descending powers of $\chi_{3y}$.
By Descartes' rule of signs, equation (\ref{eq:regionb}) has one positive real root. 
One can similarly verify that equation (\ref{eq:regionb}) has no negative real roots\footnote{Physically, this is because $m_3$ is closer to $m_+$ than $m_-$ for $X_{3y} < 0$. Hence, $m_3$ experiences a larger gravitational pull from $m_+$ into the plane of the primary and secondary orbits than the push by $m_-$ away from the plane for $m_+ > \abs{m_-}$.}.
Since equation (\ref{eq:F3}) is symmetric about the $z$-plane, there are two Lagrange points in region (b) with the same $X_{3y} > 0$.
We label these as $L_{1, \rm out}$ and $L_{2, \rm out}$ to symbolize their out-of-plane characteristics.

\subsection{Approximate locations of \texorpdfstring{$L_{1, {\rm out}}$}{L1out}, \texorpdfstring{$L_{2, {\rm out}}$}{L2out} and \texorpdfstring{$L_{3, {\rm in}}$}{L3in} for \texorpdfstring{$m_+ \gg \abs{m_-}$}{m+>>|m-|}}
\label{appendix:A3}
In this section, we study the approximate locations of $L_{1, {\rm out}}$, $L_{2, {\rm out}}$ and $L_{3, {\rm in}}$ in the $m_+ \gg \abs{m_-}$ regime.
Equation (\ref{eq:regionc2}) and (\ref{eq:regionb}) are quintics for $L_{3, {\rm in}}$, and $L_{1, {\rm out}}$ and $L_{2, {\rm out}}$, respectively, with no analytic solutions. 
We numerically obtain their solutions and plot them in Figure \ref{fig:2} and \ref{fig:3}. 
In the regime $m_+ \gg \abs{m_-}$, writing $m_- = -\epsilon m_+$ with $\epsilon > 0$ and analytically solving for the Lagrange points to first order in $\epsilon$, we obtain
\begin{equation}
    \mathbf{X}_3 \approx - \left(1 - \frac{17}{12}\epsilon\right) \mathbf{e}_y
\end{equation}
for $L_{3, {\rm in}}$
and 
\begin{equation}
    \mathbf{X}_3 \approx  \left(1 - \frac{3}{8}\epsilon^{2/3}\right) \mathbf{e}_y \pm \left(\epsilon^{1/3} - \frac{7}{128} \epsilon\right) \mathbf{e}_z
\end{equation}
for $L_{1, {\rm out}}$ and $L_{2, {\rm out}}$.


%

\end{document}